\documentclass[12pt]{article}
\usepackage{epsfig}
\usepackage{axodraw}

\newcommand{\xF}{x_{\mathrm{F}}}
\newcommand{\pt}{p_{\perp}}
\newcommand{\kt}{k_{\perp}}
\renewcommand{\c}{\mathrm{c}}
\renewcommand{\d}{\mathrm{d}}
\newcommand{\e}{\rm{e}}
\newcommand{\g}{\mathrm{g}}
\newcommand{\p}{\mathrm{p}}
\newcommand{\q}{\mathrm{q}}
\renewcommand{\u}{\mathrm{u}}
\newcommand{\D}{\mathrm{D}}

\newcommand{\cbar}{\overline{\mathrm{c}}}
\newcommand{\dbar}{\overline{\mathrm{d}}}
\newcommand{\qbar}{\overline{\mathrm{q}}}
\newcommand{\ubar}{\overline{\mathrm{u}}}
\newcommand{\Bbar}{\overline{\mathrm{B}}}
\newcommand{\Dbar}{\overline{\mathrm{D}}}

\newcommand{\Py}{{\sc{Pythia}}}

\newenvironment{Itemize}{\begin{list}{$\bullet$}%
{\setlength{\topsep}{0.2mm}\setlength{\partopsep}{0.2mm}%
\setlength{\itemsep}{0.2mm}\setlength{\parsep}{0.2mm}}}%
{\end{list}}
\newcounter{enumct}
\newenvironment{Enumerate}{\begin{list}{\arabic{enumct}.}%
{\usecounter{enumct}\setlength{\topsep}{0.2mm}%
\setlength{\partopsep}{0.2mm}\setlength{\itemsep}{0.2mm}%
\setlength{\parsep}{0.2mm}}}{\end{list}}
 
\newlength{\abstwidth}
\newlength{\captivewidth}
\begin{document}
 
\sloppy

\pagestyle{empty}

\begin{flushright}
LU TP 98--18 \\
September 1998
\end{flushright}
 
\vspace{\fill}

\begin{center}
{\LARGE\bf Production mechanisms of charm hadrons\\[4mm]
\bf in the string model}\\[10mm]
{\Large E. Norrbin\footnote{emanuel@thep.lu.se} and %
T. Sj\"ostrand\footnote{torbjorn@thep.lu.se}} \\[3mm]
{\it Department of Theoretical Physics,}\\[1mm]
{\it Lund University, Lund, Sweden}
\end{center}
 
\vspace{\fill}
 
\begin{center}
{\bf Abstract}\\[2ex]
\begin{minipage}{\abstwidth}
In the hadroproduction of charm in the context of string
fragmentation, the pull of a beam remnant at the other end of a string 
may give a charm hadron more energy than the perturbatively produced
charm quark. The collapse of a low-mass string to a single hadron is the 
extreme case in this direction, and gives rise to asymmetries between
charm and anticharm hadron spectra. We study these phenomena, and 
develop models that describe the characteristics not only of the charm 
hadrons but also of the associated event.

\end{minipage}
\end{center}
 
\vspace{\fill}

\clearpage
\pagestyle{plain}
\setcounter{page}{1} 

Asymmetries in the production spectra of charm and anticharm hadrons
(or generally, asymmetries between leading and non-leading particles)
have been observed since long \cite{oldobs}. But it is only in
recent years \cite{WA82,E769,E791} that the precision has improved 
so as to allow more detailed studies, especially in $\pi^-\p$ events.
Perturbative QCD calculations predict only very small asymmetries  
\cite{pertcharm1,pertcharm2}, so the origin of the observed asymmetries has to be 
sought in nonperturbative physics. 

Furthermore, the observed longitudinal momentum spectra of charm mesons 
are about as hard as or, in some cases, even harder than the 
perturbatively calculated charm quark spectra. This runs counter
to naive expectations, e.g. based on experience from $\e^+\e^-$ 
annihilation experiments in the 10 GeV region, where the charm hadrons 
only keep about two thirds of the original charm quark energy
\cite{RPP}. This number ought to be relevant for the hadroproduced 
charm events, since the $Q^2$ scales involved are comparable --- 
by QCD scaling violations, the fraction slowly drops with energy.  

Several scenarios have also been proposed to explain the phenomenology
\cite{variousmodels,ABG,intrinsic}. The two most frequently used ones are probably
those of string fragmentation \cite{ABG} and intrinsic charm 
\cite{intrinsic}. In this letter we develop and study the string 
fragmentation approach. The key objective is to study the collapse 
of low-mass strings into a single hadron, which provides the main 
mechanism of flavour asymmetries in our approach. The outline of
the letter is as follows. First we classify the different production 
channels for D mesons in hadron--hadron collisions, from perturbative
and nonperturbative perspectives. Then we identify the origin of the 
charm asymmetry and compare the current model with data. The critical 
aspects of charm production are thereafter studied in more detail and 
different model variations are proposed to better understand the data.
Further details and applications are intended to appear in a future 
paper.

The production of a charm hadron can be subdivided into two steps:
first the production of a $\c\cbar$ pair, followed by the
hadronization of these quarks. We will assume that the first step of
this process is adequately described by standard perturbation theory
and conventional parton distributions, i.e. without the inclusion of
any intrinsic charm component in the proton wave function. The charm 
sea is thus assumed perturbatively calculable and peaked at small $x$ 
values. Intrinsic charm and other modifications to standard perturbative 
results may well exist at {\em some} level, but here we want to show 
that it is possible to understand existing data without invoking new 
mechanisms for the production stage. 

Also with this restriction, it is not possible to obtain unambiguous 
perturbative predictions: results are sensitive to ill-determined
parameters such as the charm mass. As a simplification, we will stay with 
lowest-order matrix elements for charm production, augmented by a 
parton-shower approximation to higher-order corrections. Given the 
uncertainties already noted, this appears adequate. 

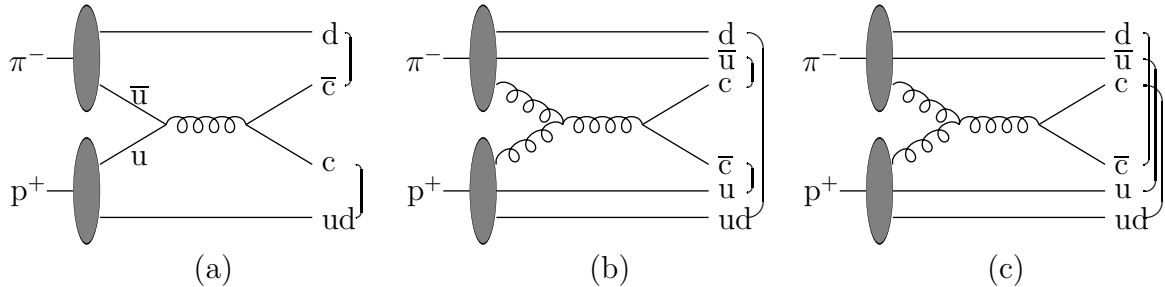
\begin{figure}
\begin{center}
\begin{picture}(450,100)(0,0)
\Text(78,0)[]{(a)}
\Text(15,30)[r]{$\p^+$}
\Text(15,80)[r]{$\pi^-$}
\Line(15,30)(25,30)
\Line(15,80)(25,80)
\GOval(30,30)(20,5)(0){0.5}
\GOval(30,80)(20,5)(0){0.5}
\Text(50,40)[b]{$\u$}
\Text(50,70)[t]{$\ubar$}
\Line(35,40)(60,55)
\Line(35,70)(60,55)
\Gluon(60,55)(90,55){3}{4}
\Line(90,55)(115,40)
\Line(90,55)(115,70)
\Text(119,40)[l]{$\c$}
\Text(119,70)[l]{$\cbar$}
\Line(35,20)(115,20)
\Line(35,90)(115,90)
\Text(119,20)[l]{$\u\d$}
\Text(119,90)[l]{$\d$}
\put(132,30){\oval(4,20)[r]}
\put(128,80){\oval(4,20)[r]}

\SetOffset(150,0)
\Text(78,0)[]{(b)}
\Text(15,30)[r]{$\p^+$}
\Text(15,80)[r]{$\pi^-$}
\Line(15,30)(25,30)
\Line(15,80)(25,80)
\GOval(30,30)(20,5)(0){0.5}
\GOval(30,80)(20,5)(0){0.5}
\Gluon(35,40)(60,55){3}{3}
\Gluon(35,70)(60,55){3}{3}
\Gluon(60,55)(90,55){3}{4}
\Line(90,55)(115,40)
\Line(90,55)(115,70)
\Text(119,40)[l]{$\cbar$}
\Text(119,70)[l]{$\c$}
\Line(35,20)(115,20)
\Line(35,90)(115,90)
\Text(119,20)[l]{$\u\d$}
\Text(119,90)[l]{$\d$}
\Line(35,30)(115,30)
\Line(35,80)(115,80)
\Text(119,30)[l]{$\u$}
\Text(119,80)[l]{$\ubar$}
\put(280,35){\oval(4,11)[r]}
\put(280,55){\oval(12,70)[r]}
\put(280,75){\oval(4,11)[r]}

\SetOffset(300,0)
\Text(78,0)[]{(c)}
\Text(15,30)[r]{$\p^+$}
\Text(15,80)[r]{$\pi^-$}
\Line(15,30)(25,30)
\Line(15,80)(25,80)
\GOval(30,30)(20,5)(0){0.5}
\GOval(30,80)(20,5)(0){0.5}
\Gluon(35,40)(60,55){3}{3}
\Gluon(35,70)(60,55){3}{3}
\Gluon(60,55)(90,55){3}{4}
\Line(90,55)(115,40)
\Line(90,55)(115,70)
\Text(119,40)[l]{$\cbar$}
\Text(119,70)[l]{$\c$}
\Line(35,20)(115,20)
\Line(35,90)(115,90)
\Text(119,20)[l]{$\u\d$}
\Text(119,90)[l]{$\d$}
\Line(35,30)(115,30)
\Line(35,80)(115,80)
\Text(119,30)[l]{$\u$}
\Text(119,80)[l]{$\ubar$}
\put(430,45){\oval(14,50)[r]}
\put(430,55){\oval(9,50)[r]}
\put(430,65){\oval(4,50)[r]}

\end{picture}
\end{center}
\caption[]{Examples of different string configurations in a $\pi^-\p$ 
collision: (a) $\u\ubar \to \c\cbar$ has a unique colour flow; (b,c)
$\g\g \to \c\cbar$ with the two possible colour flows.}
\label{fig.strings}
\end{figure}

The interesting phenomenology thus appears in the hadronization stage of
our model. Here the partons of the hard interaction and of the beam remnants 
are connected by strings, representing the confining colour field
\cite{AGIS}. Each string contains a colour triplet endpoint, a number
(possibly zero) of intermediate gluons and a colour antitriplet end.
The string topology can usually be derived from the colour flow of the
hard process. For instance, consider the process $\u\ubar \to \c\cbar$
in a $\pi^-\p$ collision. The colour of the incoming $\u$ is inherited
by the outgoing $\c$, so it will form a colour-singlet together with the
proton remnant, here represented by a colour antitriplet $\u\d$ diquark. 
In total, the event will thus contain two strings, one $\c$--$\u\d$ and
one $\cbar$--$\d$ (Fig.~\ref{fig.strings}a). In $\g\g \to \c\cbar$
a similar inspection shows that two distinct colour topologies are 
possible. Representing the 
proton remnant by a $\u$ quark and a $\u\d$ diquark (alternatively $\d$ 
plus $\u\u$), one possibility is to have three strings $\c$--$\ubar$, 
$\cbar$--$\u$ and $\d$--$\u\d$ (Fig.~\ref{fig.strings}b), and the other is the three strings 
$\c$--$\u\d$, $\cbar$--$\d$ and $\u$--$\ubar$ (Fig.~\ref{fig.strings}c).
In addition to the 
valence $\u\ubar$ annihilation and $\g\g$ fusion production mechanisms,
there are of course other possibilities, for example events involving 
sea quarks. These are included in the model but give small contributions 
to asymmetries and other observables at fixed-target energies, and therefore
will not be discussed specifically in the following.

In a process with two (or more) allowed colour flows, such as 
$\g\g \to \c\cbar$, the amplitude for either of them is calculable in
perturbation theory, but with a nonvanishing interference term between 
the two. This term, corresponding to an indeterminate colour flow,
fortunately is suppressed by a colour factor $1/N_{\mathrm{C}}^2 = 1/9$.
Its contribution to the total charm cross section therefore can be split
between the two well-determined colour flows, e.g. according to the pole 
structure of the terms \cite{hansuno}, without much resulting
ambiguity. Furthermore, we neglect the possibility of the 
$\c\cbar$ pair forming a string, either by a perturbative or 
nonperturbative colour (re)arrangement. Such mechanisms are likely
to play a significant r\^ole for $\mathrm{J}/\psi$ production, e.g., 
but should be less relevant for the open charm production to be 
considered here.

Once the string topology has been determined, the Lund string 
fragmentation model \cite{AGIS} can be applied to describe the
nonperturbative hadronization. Assuming that the 
fragmentation mechanism is universal, i.e. process-independent,
the good description of $\e^+\e^-$ annihilation data should carry over.
The main difference between $\e^+\e^-$ and hadron--hadron events is that 
the latter contain beam remnants which are colour-connected with the
hard-scattering partons. The structure of these remnants is not 
calculable from first principles, so this introduces some arbitrariness
not constrained by $\e^+\e^-$ data. In the present model these aspects 
are parameterized in beam remnant distribution functions to be considered 
later.

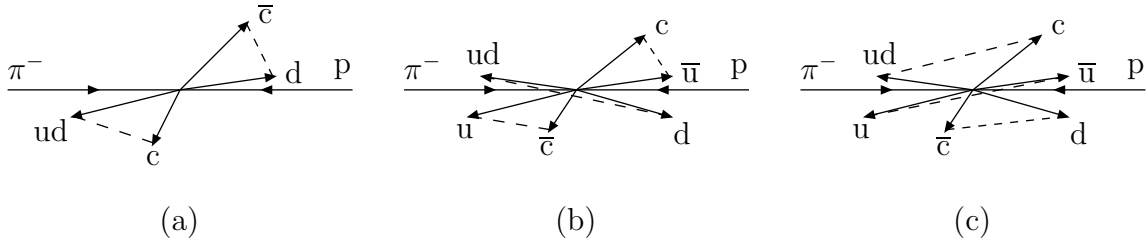
\begin{figure}\begin{center}
\begin{picture}(450,100)(0,0)
\Text(75,0)[]{(a)}
\ArrowLine(10,50)(75,50)
\Text(10,59)[l]{$\pi^-$}
\ArrowLine(140,50)(75,50)
\Text(140,58)[r]{$\p$}
\LongArrow(75,50)(110,55)
\Text(115,57)[l]{$\d$}
\LongArrow(75,50)(100,75)
\Text(105,79)[l]{$\cbar$}
\LongArrow(75,50)(35,40)
\Text(33,35)[r]{$\u\d$}
\LongArrow(75,50)(65,30)
\Text(65,27)[t]{$\c$}
\DashLine(110,55)(100,75){3}
\DashLine(35,40)(65,30){4}

\SetOffset(150,0)
\Text(75,0)[]{(b)}
\ArrowLine(10,50)(75,50)
\Text(10,59)[l]{$\pi^-$}
\ArrowLine(140,50)(75,50)
\Text(140,58)[r]{$\p$}
\LongArrow(75,50)(110,55)
\Text(115,57)[l]{$\ubar$}
\LongArrow(75,50)(100,70)
\Text(105,74)[l]{$\c$}
\LongArrow(75,50)(35,40)
\Text(33,37)[t]{$\u$}
\LongArrow(75,50)(65,35)
\Text(64,29)[]{$\cbar$}
\LongArrow(75,50)(40,55)
\Text(40,64)[]{$\u\d$}
\LongArrow(75,50)(110,40)
\Text(112,34)[l]{$\d$}

\DashLine(110,55)(100,70){3}
\DashLine(35,40)(65,35){4}
\DashLine(40,55)(110,40){4}

\SetOffset(300,0)

\Text(75,0)[]{(c)}
\ArrowLine(10,50)(75,50)
\Text(10,59)[l]{$\pi^-$}
\ArrowLine(140,50)(75,50)
\Text(140,58)[r]{$\p$}
\LongArrow(75,50)(110,55)
\Text(115,57)[l]{$\ubar$}
\LongArrow(75,50)(100,70)
\Text(105,74)[l]{$\c$}
\LongArrow(75,50)(35,40)
\Text(33,37)[t]{$\u$}
\LongArrow(75,50)(65,35)
\Text(64,29)[]{$\cbar$}
\LongArrow(75,50)(40,55)
\Text(40,64)[]{$\u\d$}
\LongArrow(75,50)(110,40)
\Text(112,34)[l]{$\d$}

\DashLine(65,35)(110,40){3}
\DashLine(35,40)(110,55){4}
\DashLine(40,55)(100,70){4}

\end{picture}\end{center}
\caption[]{Strings (dashed) in a $\pi^-\p$ collision corresponding to the colour flows
in Fig.~\ref{fig.strings}a, b and c respectively;
(a) $\u\ubar \to \c\cbar$ and (b,c) $\g\g \to \c\cbar$. If e.g. the colour-singlet
system $\c$-$\ubar$ in (b) has a small invariant mass we will call it a cluster
and hadronize it by the procedure described in the text.}
\label{fig.clusters}
\end{figure}

Depending on the invariant mass of a string, practical considerations
lead to the need to distinguish three hadronization prescriptions:
\begin{Enumerate}
\item {\em Normal string fragmentation}.
In $\e^+\e^-$ collisions the string system has a mass equal to the full 
CM energy, neglecting the not-too-frequent $\g \to \q\qbar$ shower
branchings which splits a string in two. This situation is ideal for 
an iterative fragmentation scheme, for which the assumption of a
continuum of phase-space states is essential. The average multiplicity
increases linearly with the string `length' which means, neglecting 
gluon-emission effects, logarithmically with the string mass.
In practice, this approach can be used for all strings
above some cut-off mass of a few GeV. 
\item {\em Cluster decay}.
In a hadron collision at fixed-target energies it frequently happens 
that a colour-singlet system contains two partons moving in the same 
general direction, see Fig.~\ref{fig.clusters}. This gives
the system a small invariant mass, for which maybe only two-body
final states are kinematically accessible. The continuum assumption
above then is not valid, and the traditional iterative Lund scheme is 
not applicable. We call such a low-mass string a cluster, and treat
it differently from above. When kinematically possible, a $\c$--$\qbar$ 
cluster will decay into a charmed meson and a light meson by the 
production of a light quark--antiquark pair in the colour force field 
between the two cluster endpoints, with the new quark flavour selected 
according to the same rules as in normal string fragmentation. The decay 
kinematics is for the time being assumed to be isotropic in the 
rest frame of the cluster, which is the behaviour to be expected  
in the limit of vanishing phase space.
\item {\em Cluster collapse}.
This is the extreme case of the above situation, where the string 
mass is so small that the cluster cannot decay into two hadrons.
It is then assumed to collapse directly into a hadron resonance,
inheriting the flavour content of the string endpoints. The original 
continuum of string/cluster masses is replaced by a discrete set
of hadron masses, mainly $\D$ and $\D^*$. By local duality arguments
\cite{duality} we assume that this does not change the total rate 
of charm production. This is related to the argument used in the 
$\e^+\e^- \to \c\cbar$ channel, that the cross section in
the $\mathrm{J}/\psi$ and $\psi'$ peaks is approximately equal to a
purely perturbatively calculated $\c\cbar$ production cross section
restricted to the below-$\D\Dbar$-threshold region. Similar
relations have also been studied e.g. for $\tau$ decay to hadrons
\cite{taudecay}, and there shown to be valid to good accuracy.
In the current case, the presence of other strings in the 
event additionally allows soft-gluon exchanges to modify 
parton momenta as required to obtain correct hadron masses.
Traditional factorization of short- and long-distance physics would 
then also protect the charm cross section. Local duality and factorization
do not specify how to conserve the overall energy and momentum of an 
event, when a continuum of $\cbar\d$ masses is to be replaced by a 
discrete $\D^-$ one, however. This will therefore be one 
of the key points to be studied below.
\end{Enumerate}
Nature will not be so crude in its classification as we have been here,
but hopefully our ansatz should be close enough to give a good 
first approximation.

\begin{figure}[t]
\begin{center}
\mbox{\epsfig{file=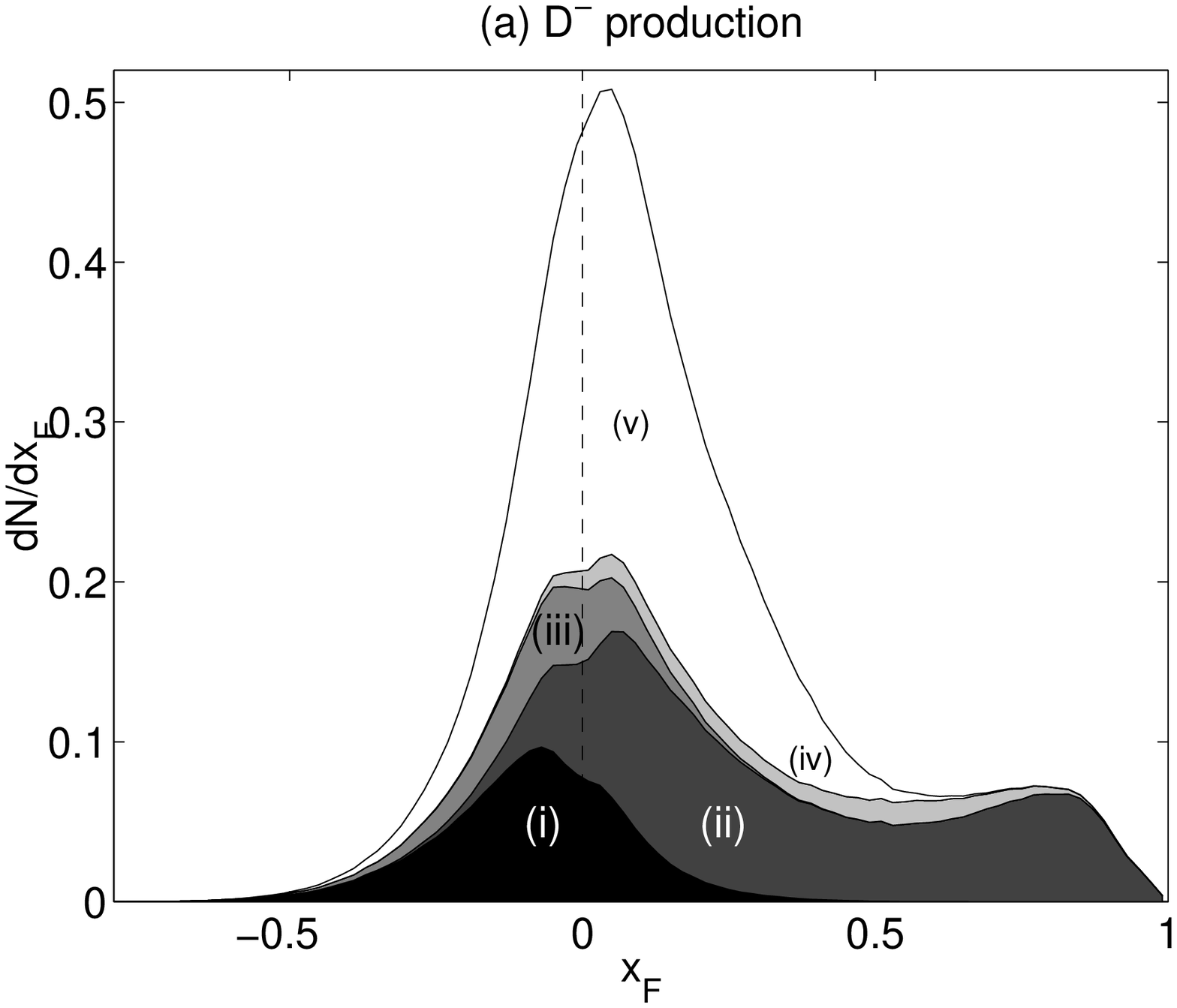, width=79mm}}
\mbox{\epsfig{file=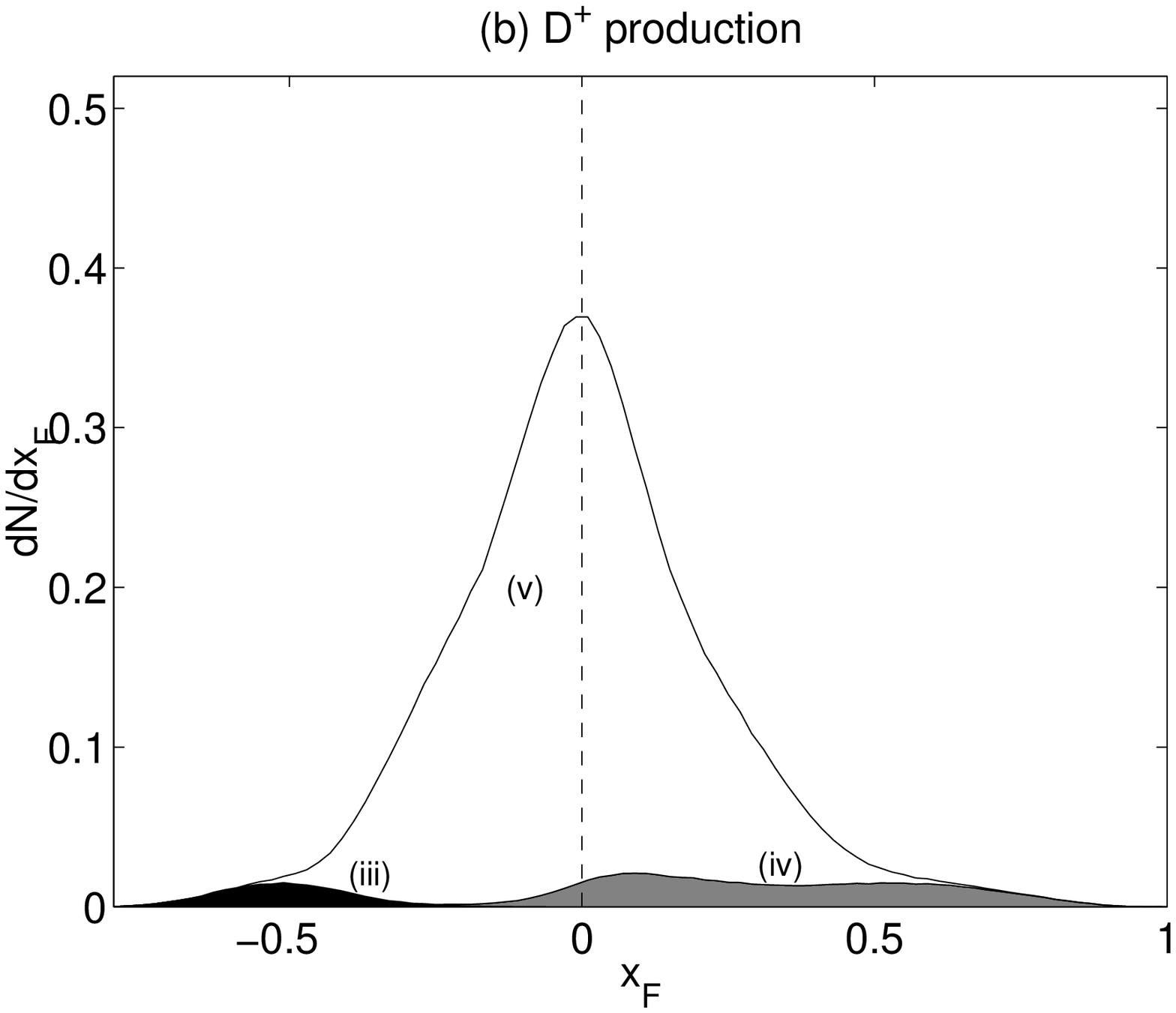, width=79mm}}
\mbox{\epsfig{file=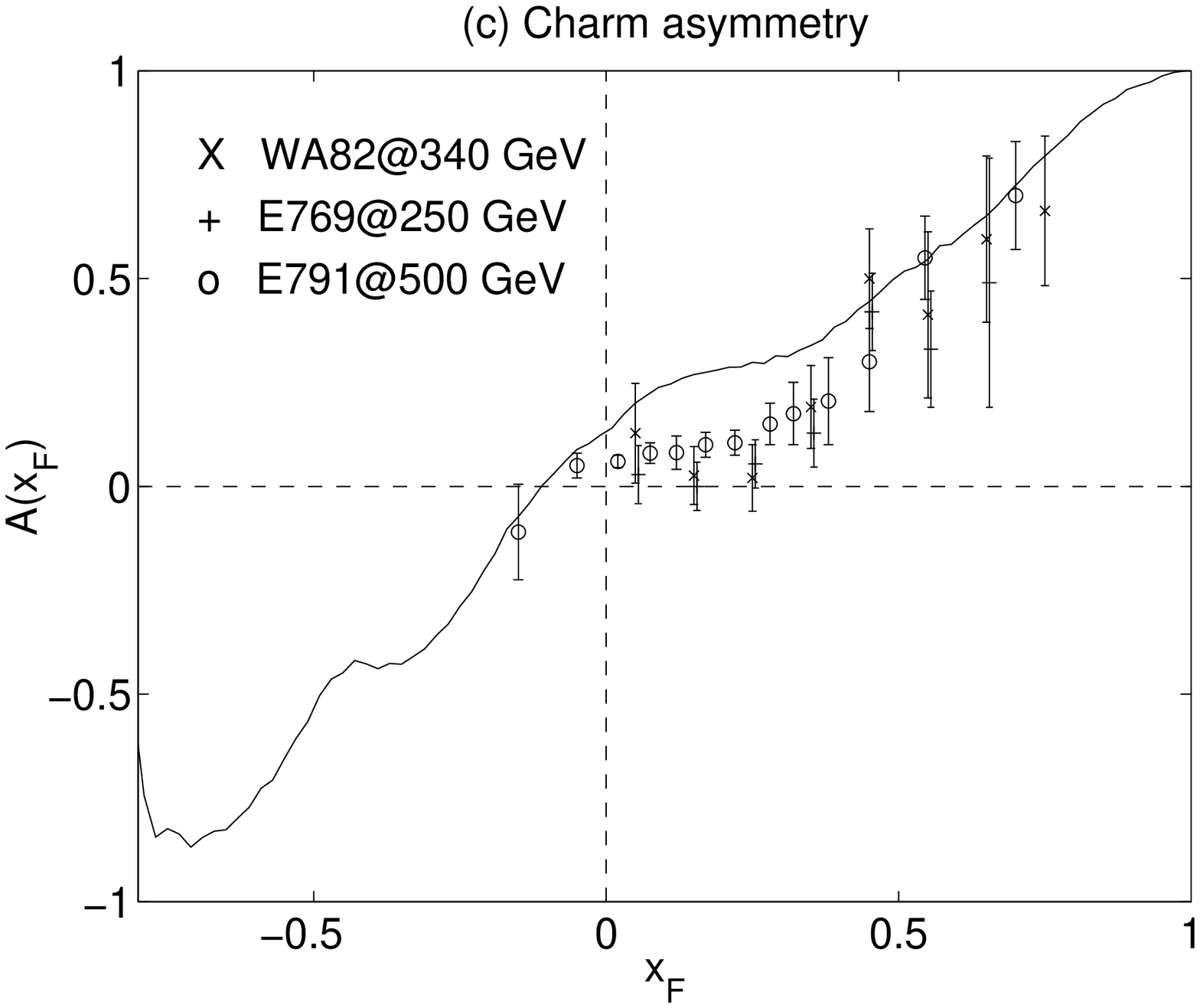, width=79mm}}
\end{center}
\caption[]{
$\D^{\mp}$ meson production in a $\pi^-\p$ collision at a $\pi^-$ beam momentum
of 500 GeV, using the default Lund Model.
$\xF$ distribution (normalized per $\c\cbar$ event) of (a) $\D^-$ and (b) $\D^+$
for different production channels:
(i) Cluster collapse, light quark from p end,
(ii) Cluster collapse, light quark from $\pi^-$ end,
(iii) Cluster decay, light quark from p end,
(iv) Cluster decay, light quark from $\pi^-$ end and
(v) String fragmentation.
(c) The resulting asymmetry, Eq~(\ref{eq.asymmetry}). Also shown is data from
\cite{WA82,E769,E791}.}
\label{fig.default}
\end{figure}

Charm cross sections are often presented as functions of Feynman-$x$, 
$\xF = p_{\mathrm{L}} / p_{\mathrm{L,max}} \approx %
2 p_{\mathrm{L}}/E_{\mathrm{CM}}$, i.e. the longitudinal 
momentum fraction of the meson in the rest frame of the event
\cite{WA82,E769,E791}. The $\xF$ distribution for the different
production channels in the standard Lund model is shown in 
Fig.~\ref{fig.default}, both for $\D^-$ and $\D^+$. 
We use \Py~6.1 \cite{Pythia} with standard
parameters to generate the plots.

In perturbative QCD the $\xF$ spectra of produced charm/anticharm 
quarks are identical to leading order, and the effects of higher 
orders are very small in this respect \cite{pertcharm1,pertcharm2}. Therefore any 
asymmetry between charm and anticharm hadrons is a simple measure of
nonperturbative effects. In $\pi^-\p$ experiments the charm asymmetry 
is traditionally defined as
\begin{equation}
A = A(\xF,\pt)=
\frac{\sigma(\D^-)-\sigma(\D^+)}{\sigma(\D^-)+\sigma(\D^+)} ~,
 \label{eq.asymmetry}
\end{equation}
which measures the relative abundance of $\D^-$ over $\D^+$
as a function of $\xF$ and $\pt$. We will here mainly consider
the variation with $\xF$, integrated over $\pt$, since $A$
does not vary much with $\pt$ over the experimental range of $\xF$ values
\cite{E769,E791}. The asymmetry in \Py~is compared
to experiment in \cite{WA82,E769,E791}, and the model shows general 
agreement with experiment, except in the region $0 \leq \xF \leq 0.4$ 
where the model is predicting a larger asymmetry than found in the data.
One collaboration \cite{E791} has tuned the parameters
of \Py~to obtain a good description of data also in this region,
see below.

In our model the main source of asymmetry for all $\xF > 0$
is the production of $\D^-$ ($\cbar\d$) mesons via cluster collapse
involving a valence $\d$ from the pion beam, Fig.~\ref{fig.default}a.
This production channel is not open for $\D^+$  ($\c\dbar$) since the
$\pi^-$ does not contain a valence $\dbar$. A number of model 
aspects influence the rate and character of cluster collapses,
and in the following we will review the main ones.

The {\em charm mass} enters in the perturbative matrix element as well
as in the phase space, and so has a strong influence on the total 
charm cross section, but we have checked that the leading-order $\xF$ distribution
is not affected much. It also, together with the {\it light-quark masses},
sets the threshold of the charm-cluster mass spectrum.
The $\c$ mass to use need not even be the same in the two applications,
cf. the familiar distinction between current algebra and constituent 
masses. A large combined mass of the charm and light quarks in a cluster
dramatically reduces the importance of the cluster collapse mechanism.

\begin{figure}[t]
\begin{center}
\mbox{\epsfig{file=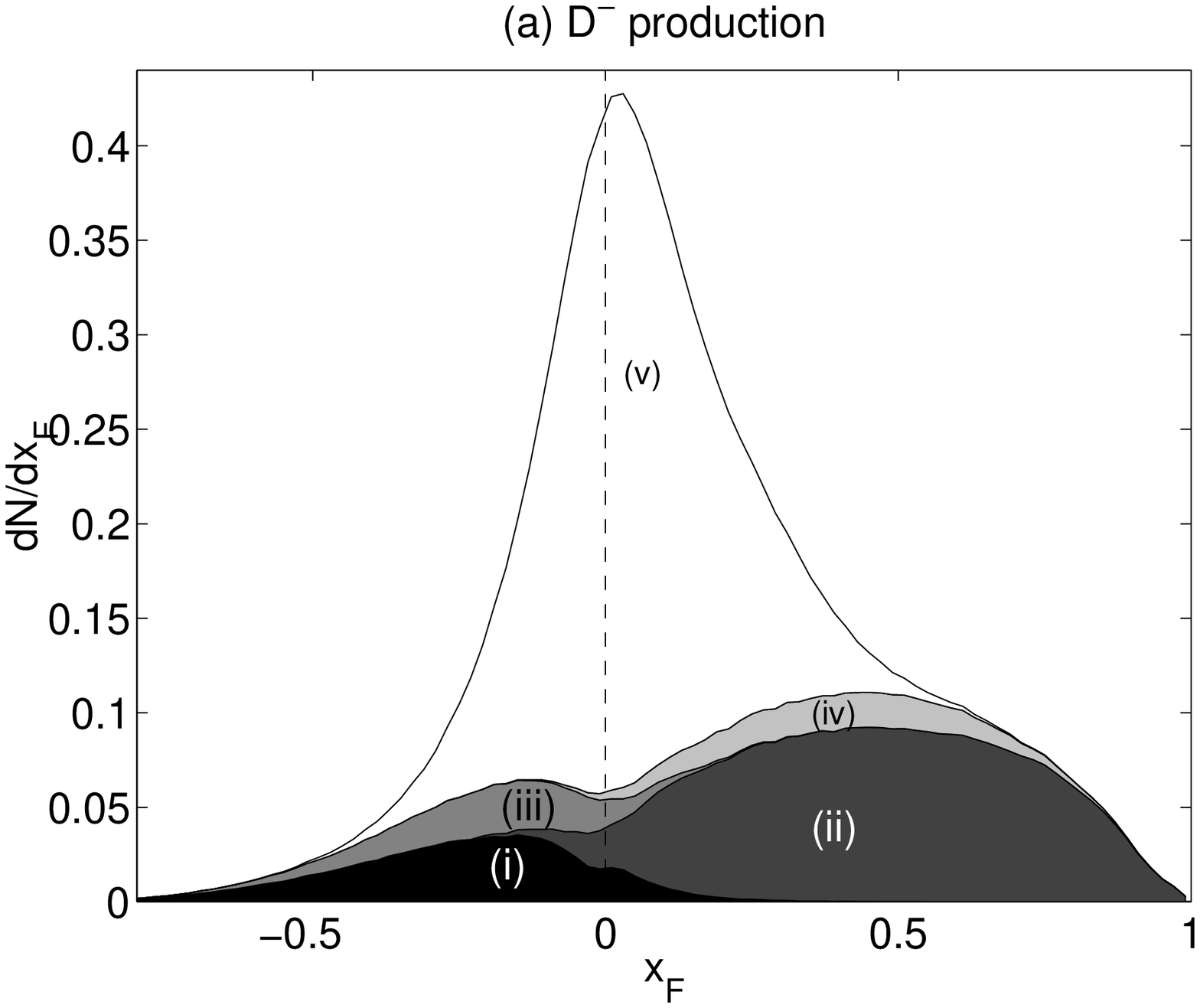, width=79mm}}
\mbox{\epsfig{file=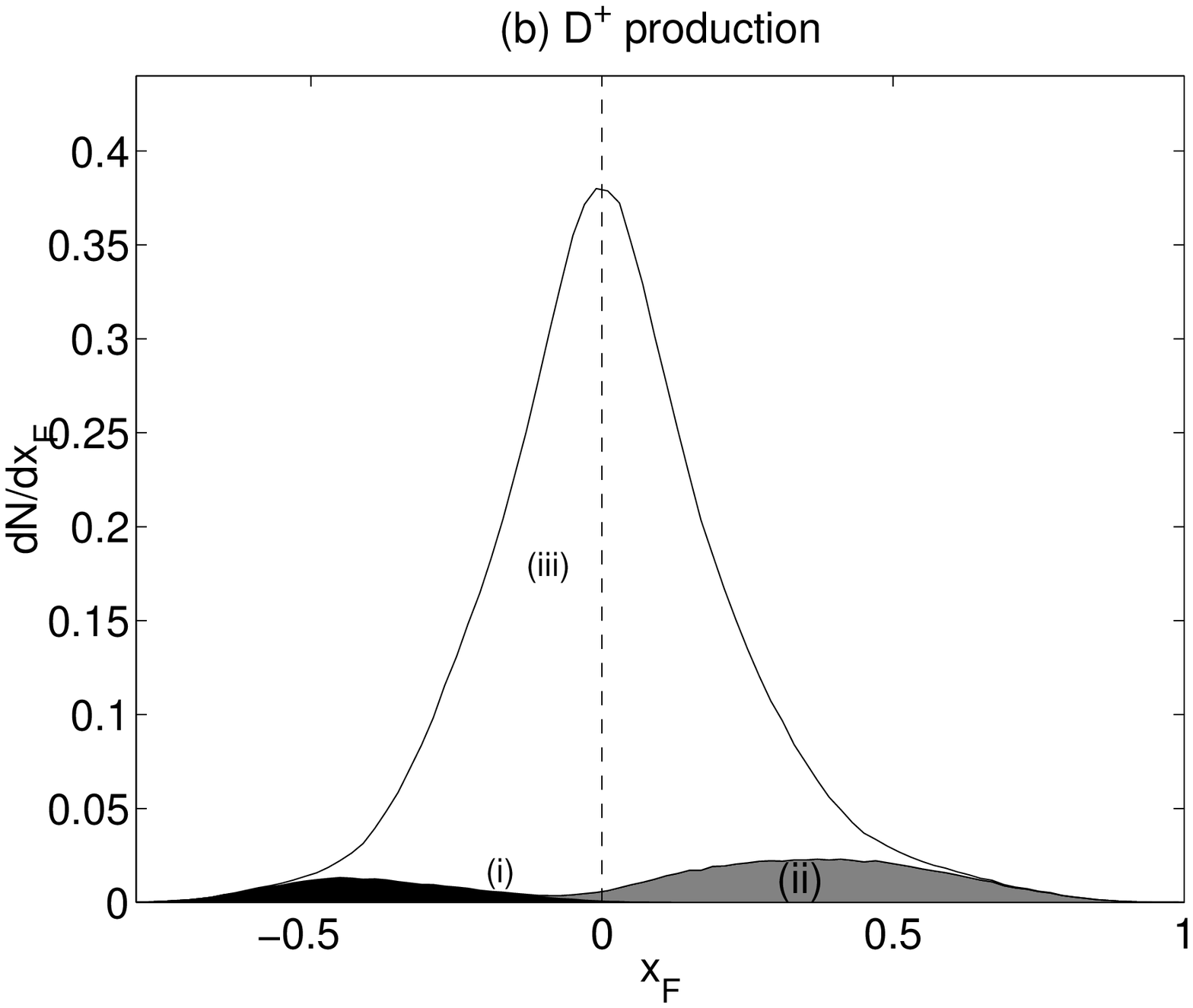, width=79mm}}
\mbox{\epsfig{file=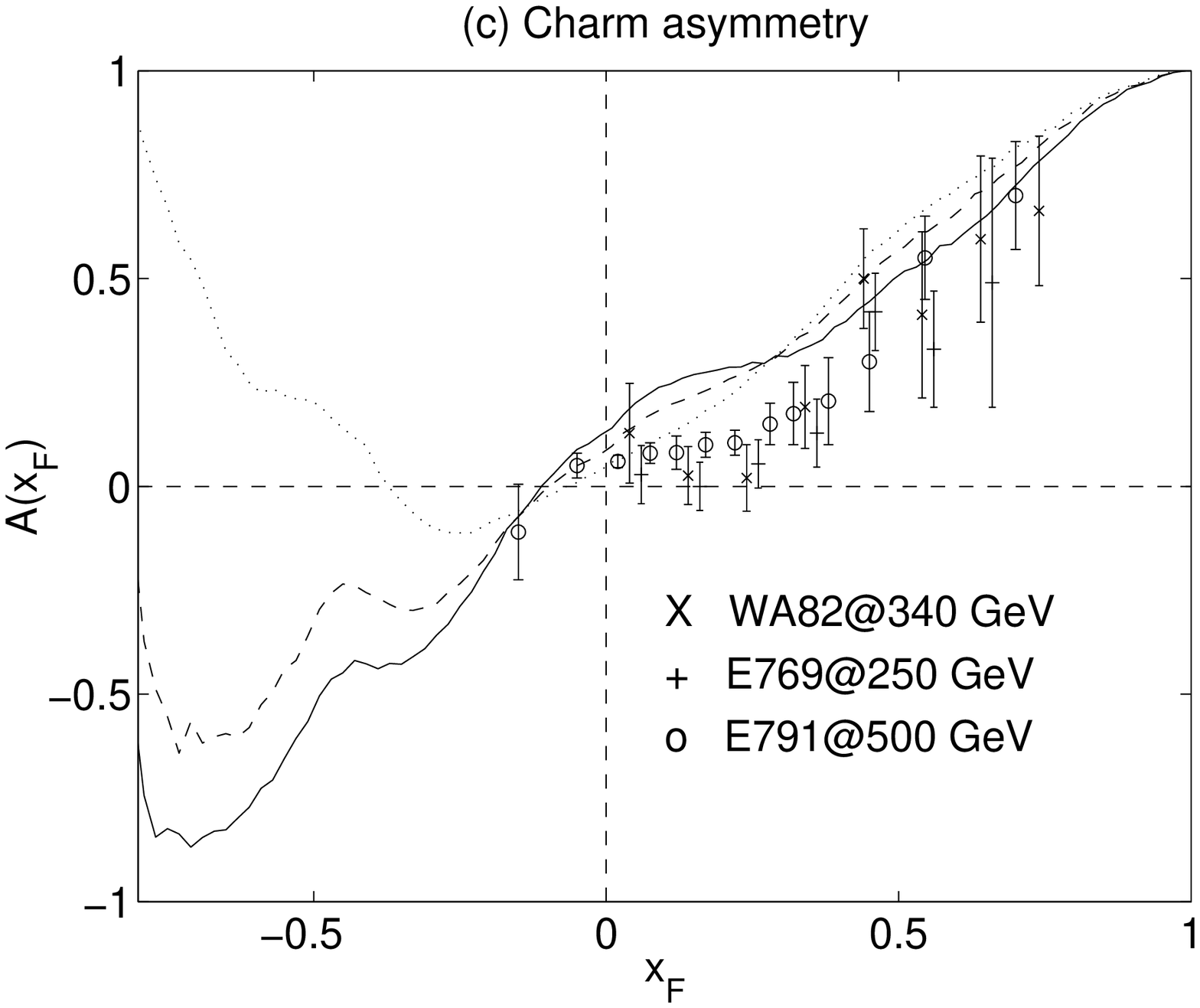, width=79mm}}
\end{center}
\caption[]{$\xF$ distribution of (a) $\D^-$ and (b) $\D^+$ for different
production channels (cf. Fig.~\ref{fig.default}, which corresponds to the default of
extreme uneven sharing) in a $\pi^-\p$ collision
at a $\pi^-$ beam momentum of 500 GeV using an even sharing of energy between the quarks
in a beam remnant.
(c) the resulting asymmetry using three different choices of BRDF's:
$\propto (1-\chi)^k/\sqrt{\chi^{2}+c_{\mathrm{min}}^2}$,
$\propto (1-\chi)^k/\sqrt[4]{\chi^{2}+c_{\mathrm{min}}^2}$ and 
$\propto (1-\chi)^l$ respectively, with $(k,l)=(1,0)$ for the pion 
and $(3,1)$ for the proton remnant, and $c_{\mathrm{min}}=0.6$~GeV/$\mathrm{E}_{\mathrm{cm}}$.
They correspond to an uneven (full), intermediate (dashed) and even
(dotted) sharing of energy. For the $\pi^-$ the three cases correspond
to $<\chi>$ = 0.14, 0.23 and 0.5 respectively. Also shown is data from
\cite{WA82,E769,E791}.}
\label{fig.brdf}
\end{figure}

In the valence $\u\ubar$ annihilation mechanism, the $\pi^-$ beam remnant 
is a $\d$ quark, so no internal structure need be specified.
In the $\g\g$ fusion process, however, the remnant is a $\ubar\d$ system
in a colour octet state, i.e. attached to two strings. A convenient
approach is to imagine this system split into two separate $\ubar$ and
$\d$ string endpoints. The {\em beam remnant distribution function} (BRDF)
is introduced to describe how the (light-cone) momentum of the remnant
is shared between the two, in fractions $\chi$ and $1 - \chi$, 
respectively. For an octet meson remnant the $\chi$ distribution is
always implicitly symmetrized between the $\q$ and $\qbar$, while for
an octet baryon remnant one quark (picked at random among the three)
takes the fraction $\chi$ and the remaining colour antitriplet diquark
$1 - \chi$. To study the dependence of the asymmetry on the BRDF, we 
will consider two extreme cases and one intermediate. In one extreme 
one quark tends to take a small fraction of the available energy,
much like the parton distributions in a hadron. In the other extreme, 
naive counting rules are used and the energy is, on the average, shared 
evenly between the quarks in the remnant. In Fig.~\ref{fig.brdf}
the individual spectra as well as the asymmetry is shown for the full $\xF$ range.

It is interesting to note the difference between the regions $\xF<0$
and $\xF>0$. For $\xF>0$ the $\D^-$ is a leading particle
and the asymmetry is attributed to cluster collapses involving a
d-quark from the pion beam. In the proton fragmentation region ($\xF<0$)
the $\D^-$ is still leading and $\D^+$ non-leading, so naively you
would expect the asymmetry to be positive.
However, when using an uneven sharing of energy in the proton
beam remnant, diquark effects become prominent. As is seen in
Fig.~\ref{fig.clusters} the c is always connected to diquarks
and the $\cbar$ is connected to quarks. Therefore $\D^+$ mesons produced
from a cluster (containing a diquark) via cluster decay will,
on the average, be harder than $\D^-$ from cluster collapses, cf.
Fig.~\ref{fig.default}. We see that there are two competing effects for $\xF<0$,
one that favours $\D^+$ (diquark effect), and one that favours $\D^-$
(cluster collapse). The strength of the diquark effect
depends strongly on the assumed energy sharing in the proton beam remnant.
This is seen clearly in Fig.~\ref{fig.brdf}c where an even sharing
makes the asymmetry change sign for $\xF<0$.
If this region could be examined
experimentally ($\p$ beam on $\p$ target) it would give us a hint 
at which distribution should be used. Possibly one could
use data from other (related) experiments
to assess a reasonable choice of BRDF.

The partons entering the hard interaction are traditionally taken to
have a nonvanishing {\em primordial $\kt$}, seen as a purely nonperturbative 
Fermi motion of partons inside the incoming hadrons, in addition to 
whatever is provided by perturbative gluon emission. Typical values
should thus be 300--400 MeV. In many connections, also for charm
in the current energy range \cite{pertcharm2}, it has been 
noted that much higher values are required, at or above 1 GeV.
This remains somewhat of a mystery, which we do not attempt to solve 
here. The choice of primordial $\kt$ distribution is of non-negligible
importance, both by providing a $\pt$ kick to the produced charm
quarks and, by momentum conservation, an opposite kick to the beam
remnants.  

\begin{figure}
\begin{center}
\mbox{\epsfig{file=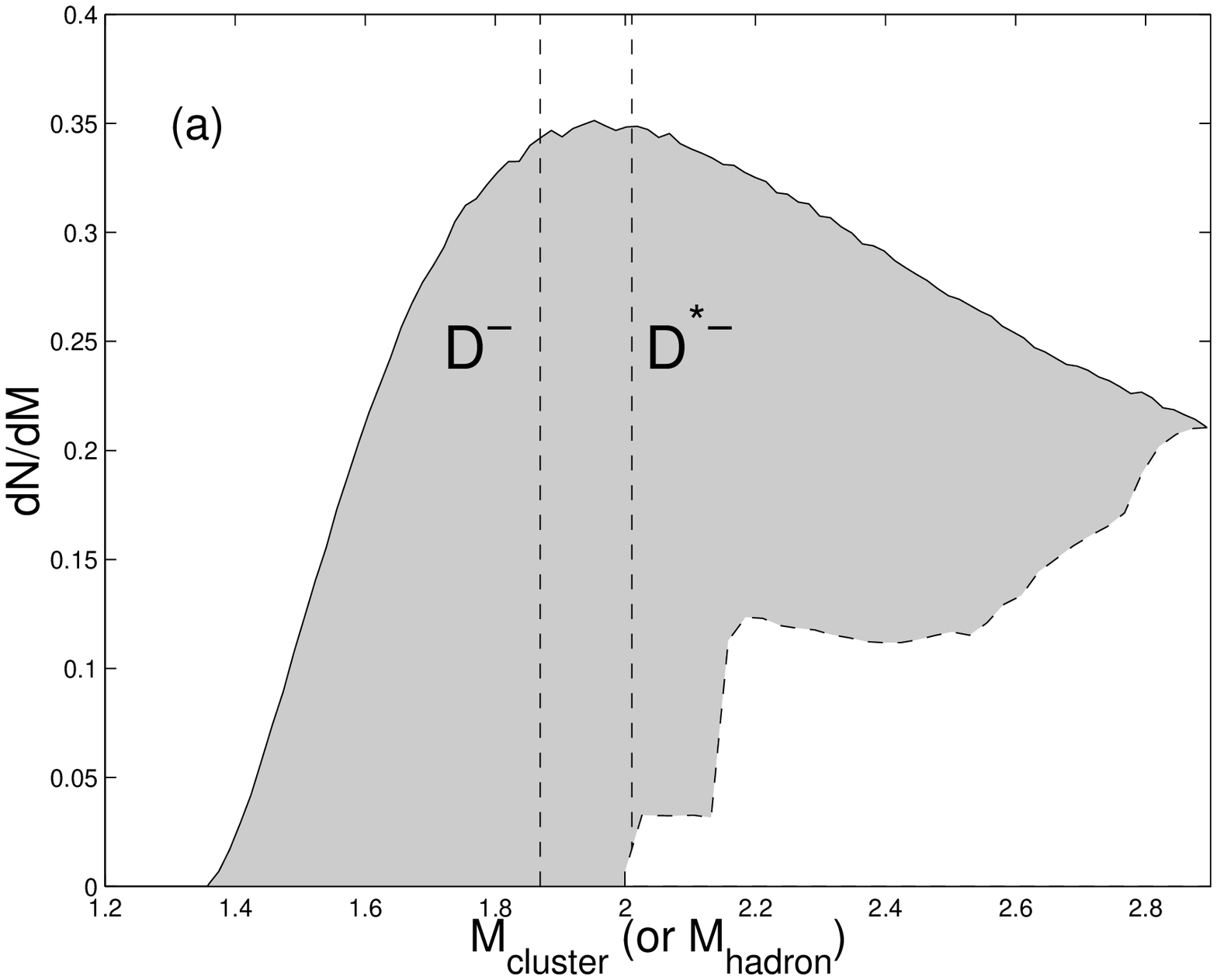,width=79mm}}
\mbox{\epsfig{file=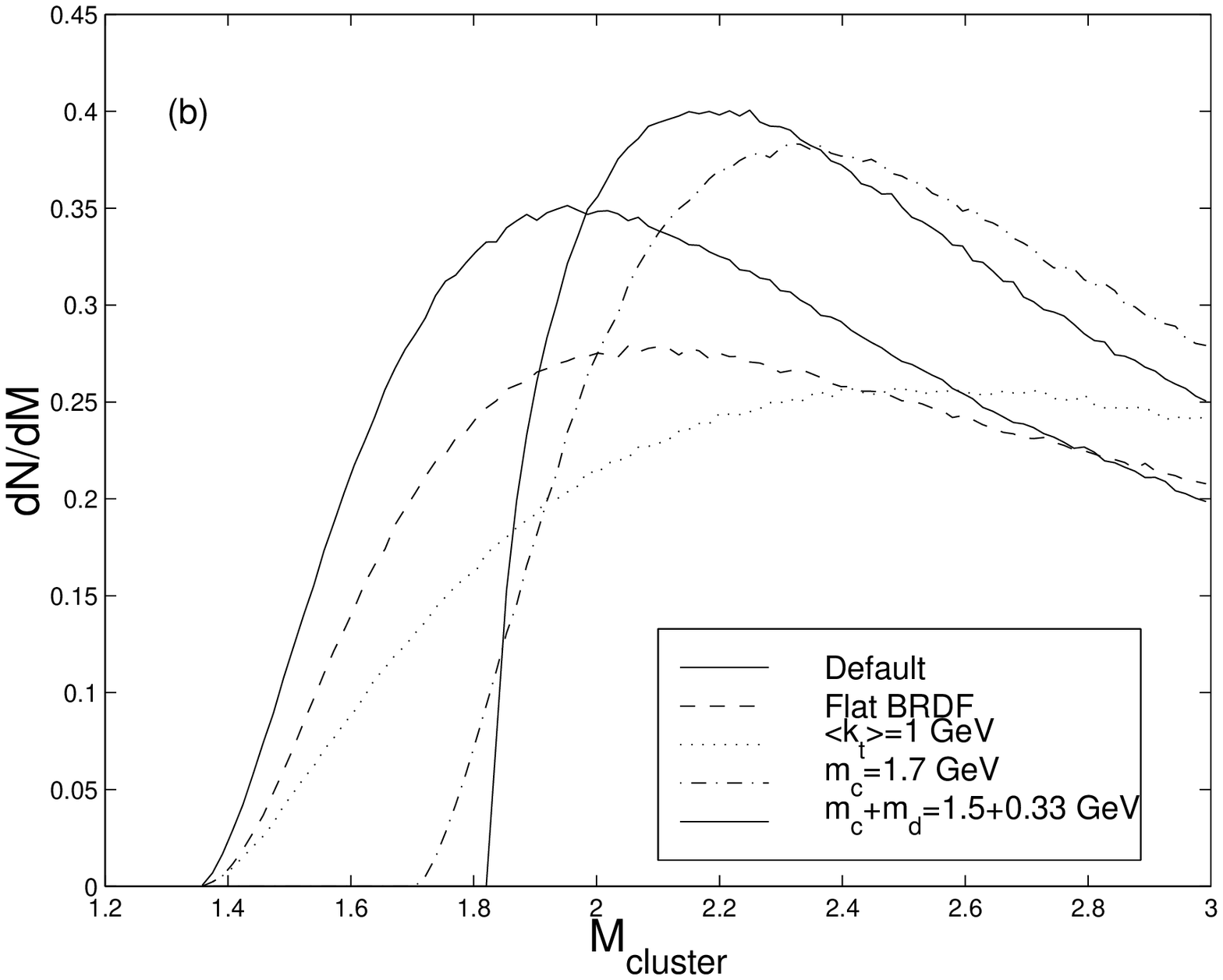,width=79mm}}
\end{center}
\caption[]{(a) Distribution of cluster (full) and meson (dashed) masses in the
string model. Clusters within the gray area collapse to $\D^-$ or $\D^{*-}$.
(b) Dependence of the parton-level mass distribution on some parameters of the model.}
\label{fig.cmass}
\end{figure}

The cluster mass spectrum is affected both by the choice of charm 
mass, of BRDF's and of primordial $\kt$ distributions. 
As a typical example, the $\cbar$--$\d$ singlet produced
in Fig.~\ref{fig.clusters}a  can have the mass distribution of 
Fig.~\ref{fig.cmass}a when calculated on the parton-level with
$\mathrm{m}_c=1.35$ GeV and $\mathrm{m}_d \approx \mathrm{m}_u \approx 0$,
or as in Fig.~\ref{fig.cmass}b if some parameters are varied.
But we know that the observed mass spectrum of produced particles 
consists of peaks at the $\D^-$ and $\D^{*-}$ masses and then a 
continuum above the $\D^- \pi^{0}$ threshold, cf. Fig.~\ref{fig.cmass}a.

The onset of the continuum depends on the assumed
{\em threshold behaviour}. In one extreme, a $\D\pi$
pair is always formed when the cluster mass allows it.
In another, this continuum state is only reached after
a succession of thresholds: $\D\pi$, $\D^*\pi$,
$\D\rho$, $\D^*\rho$, etc. Thus a fraction of events
may collapse to a single resonance also above threshold,
cf. the gray area in Fig.~\ref{fig.cmass}a.
In the default implementation, exactly one attempt is
made to form two hadrons by a random choice of allowed
flavours and spins (using the string fragmentation
relative probabilities), and is allowed to succeed
if the sum of hadron masses is below the cluster mass.
By increasing the number of attempts made before
giving up, the behaviour interpolates to the first
extreme above. Thus we see that the mass spectrum
of collapsing clusters is affected in the low end 
by quark masses and in the high end by the transition
to two-body states, altogether giving a large range of
possible cluster collapse rates.

When a collapse occurs, {\em confinement effects} have to
project the continuum of string masses onto the observed
hadron mass spectrum. Because of the aforementioned local duality and
factorization arguments, the total area of the spectrum should 
be conserved in the process. How the projection should be done is not 
known from first principles, however. 

One conceivable strategy could be to introduce a weight function 
consisting of $\delta$ function peaks at the $\D^-$ and $\D^{*-}$ masses, 
with suitably adjusted normalizations, and then a step function at the 
$\D^- \pi^{0}$ threshold. This weight function, when multiplied with
the partonic mass spectrum, should then give the hadron-level mass 
spectrum. Such an approach is not well suited for
Monte Carlo simulation, since the string mass is a complicated 
function of a number of variables and therefore the $\delta$ function
cannot easily be integrated out. However, on general grounds,
we do not expect the overall distribution of event 
characteristics to differ significantly between events with a
$\cbar$--$\d$ string mass exactly equal to the $\D^-$ one, and 
events where the string mass is maybe 100~MeV off. An appealing
shortcut therefore is to accept all partonic configurations  
and thereafter introduce some `minimal' adjustments to the 
kinematics to allow hadrons to be produced on the mass shell.
Such a strategy would be consistent not only with local duality arguments,
but also with the presence of soft final-state interactions,
i.e. the exchange of nonperturbative gluons that can carry some
amount of momentum between the low-mass string and the surrounding
hadronic system. In the following we will therefore adopt the 
language of `gluons' transferring energy and momentum between the
strings in a collision, while leaving unanswered the question on the
exact nature of those `gluons'. Specifically, we will not address the 
possibility of changes in the colour structure of events by such 
`gluons'.

In this letter we will consider two different choices of energy 
shuffling schemes that can be said to be of opposite nature. 
We will further show that, as far as the asymmetry is concerned, 
the observable differences are small.

The first approach is the standard one in \Py. It consists of shuffling 
momentum to the parton $i$ in the event that has the largest invariant 
mass when combined with the cluster cl that is supposed to collapse to a
$\D$ (say). The new $i$ and $\D$ momenta are given by
\begin{eqnarray}
 p_{\D} & = & (1 - \delta) p_{\mathrm{cl}} + \epsilon p_i ~, \nonumber \\
 p'_i   & = & (1 - \epsilon) p_i + \delta p_{\mathrm{cl}} ~,
\end{eqnarray}
where $\epsilon$ and $\delta$ are determined by the conditions
$p_{\D}^2 = m_{\D}^2$ and $p'^2_i = p_i^2 = m_i^2$. Using the parton
furthest away from the cluster has the advantage of minimizing the 
required momentum transfer, but does not offer a particularly 
appealing picture physically. In the following we therefore try 
to formulate a scheme where energy and momentum is shuffled to 
partons or strings in the vicinity of the cluster.

This problem can be approached in several ways, and we want to mention 
a few. The basic idea is to emit a soft gluon from the cluster, 
taking away the energy and momentum needed to put the remainder 
on the hadronic mass shell. One way to do this is simply to rescale the 
four-momentum of the cluster in order to give it the correct 
invariant mass. The gluon can then have both positive and negative
masses and energies, which may be conceptually unappealing but
not forbidden in principle. Worse is that such a rescaling can give
$\xF > 1$, i.e. kinematical inconsistencies. Another approach is to 
let the cluster decay isotropically into a hadron and a massless 
gluon. The gluon can still have both positive and negative 
energies, but this is not a problem in itself since the gluon is connected 
to the nearest string piece, which is then hadronized according to the 
standard string-with-gluons scheme \cite{stringwithg}. A gluon with 
positive energy will increase the energy and mass of the string 
while a gluon with negative energy will decrease it. It is the string 
which is the physical entity, and it of course has to have
a positive energy and mass. The only observable effect of the extra
gluon is a slight increase (decrease) in the average multiplicity 
in the phase space neighbourhood of the inserted  
positive (negative) energy gluon, but this effect is too small to 
be seen in an actual experiment. Even so, there are problems with the 
simple approach. For instance, if a negative-energy gluon is
connected to a string that already has a small mass, it can give the 
string a negative mass and this is not permitted. So, if there are no 
sufficiently massive strings left in the event, the gluon cannot be 
connected anywhere, and this method cannot be used. Only if the radiated 
gluon has positive energy can the mechanism always be used without 
problems.

\begin{figure}
\begin{center}
\mbox{\epsfig{file=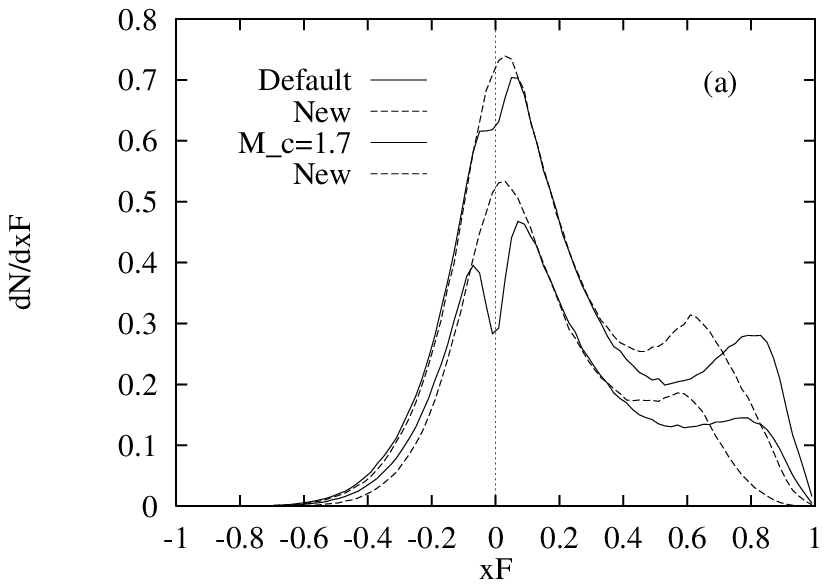,width=79mm}}
\mbox{\epsfig{file=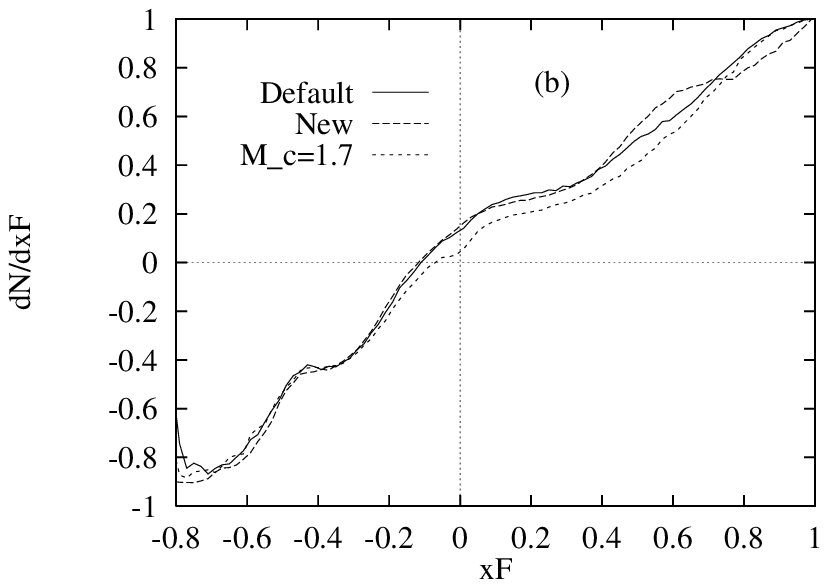,width=79mm}}
\end{center}
\caption[]{Some results for different cluster collapse algorithms:
(a) The momentum spectrum for $\D^-$ mesons produced from clusters 
(normalized to the total $\D^-$). The two
top curves are with default parameters and the lower ones are with 
an increased charm mass (from 1.35 to 1.7 GeV).
(b) the resulting asymmetry.}
\label{fig.collapse}
\end{figure}

In practice, therefore, the negative-gluon approach is too 
error-prone. When the cluster mass is smaller than the intended 
hadron mass another approach is instead used to simulate
the same effect, i.e. to take up energy from the vicinity 
of the cluster: a nearby string is allowed to emit a gluon
which, when absorbed by the cluster, gives it the right mass. 
In more detail, imagine a string with endpoints $\q$ and $\qbar$.
Form a weighted sum of the endpoint momenta
\begin{equation}
  p_s = \alpha p_{\q} + \beta p_{\qbar} 
  = \frac{p_{\qbar} p_{\mathrm{cl}}}{p_{\q} p_{\qbar}} p_{\q}     
  + \frac{p_{\q} p_{\mathrm{cl}}}{p_{\q} p_{\qbar}} p_{\qbar} ~,     
\end{equation}
so that the end of the string that is closest to the cluster
is weighted up relative to the one further away. Thereafter
define
\begin{equation}
 p_{\D} = p_{\mathrm{cl}} + \delta p_s ~,
\end{equation}
with $\delta$ determined by the constraint $p_{\D}^2 = m_{\D}^2$.
The meson will then have the correct mass and the string endpoint
momenta are scaled down by factors $1 - \delta \alpha$ and 
$1- \delta \beta$, respectively. Special cases need to be introduced
to avoid e.g. $1 - \delta \alpha < 0$, but these affect only a small
fraction of the events, and can be left aside here.  

In summary, the new algorithm transfers a `gluon' from the cluster
to the nearest string if the original cluster mass is above the
$\D$ meson mass, and transfers a `gluon' the other way if the mass
instead is below. Some results are shown in Fig.~\ref{fig.collapse}. 
As can be seen, the largest difference is in the distribution of the 
collapsed mesons at high $\xF$, where the spectrum is somewhat 
softer in the new scheme than in the old one. Another difference is 
around $\xF=0$, where the old algorithm tends to push the meson away 
from the middle, creating an unphysical dip in the distribution.
This effect is seen more clearly in the distributions for a larger 
charm mass. The distribution with $m_{\c} = 1.7$~GeV is also included 
to show the relative importance of the collapse mechanism on the 
asymmetry. Clearly the charm mass --- by regulating the amount
of collapses --- provides a much larger uncertainty (except possibly
at very large $\xF$) than the choice of collapse scheme but, if nothing else,
the new algorithm remedies some cosmetic problems of the old one.
The new algorithm also changes some other characteristics of the event, e.g. the
mean multiplicity is changed by $\approx 2$ \%, but this too is
a small effect compared to the other uncertainties that we have mentioned.

\begin{figure}
\begin{center}
\mbox{\epsfig{file=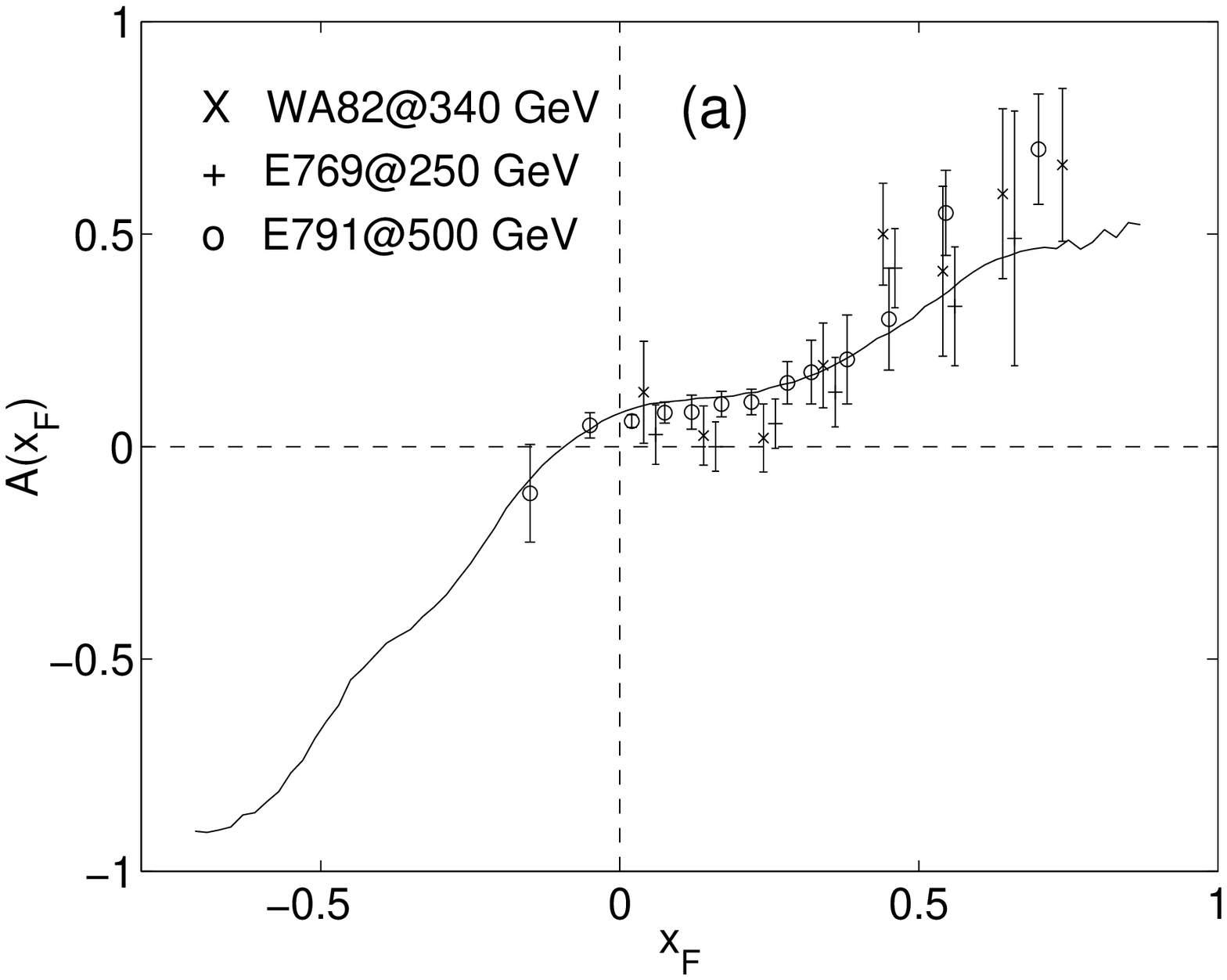,width=79mm}}
\mbox{\epsfig{file=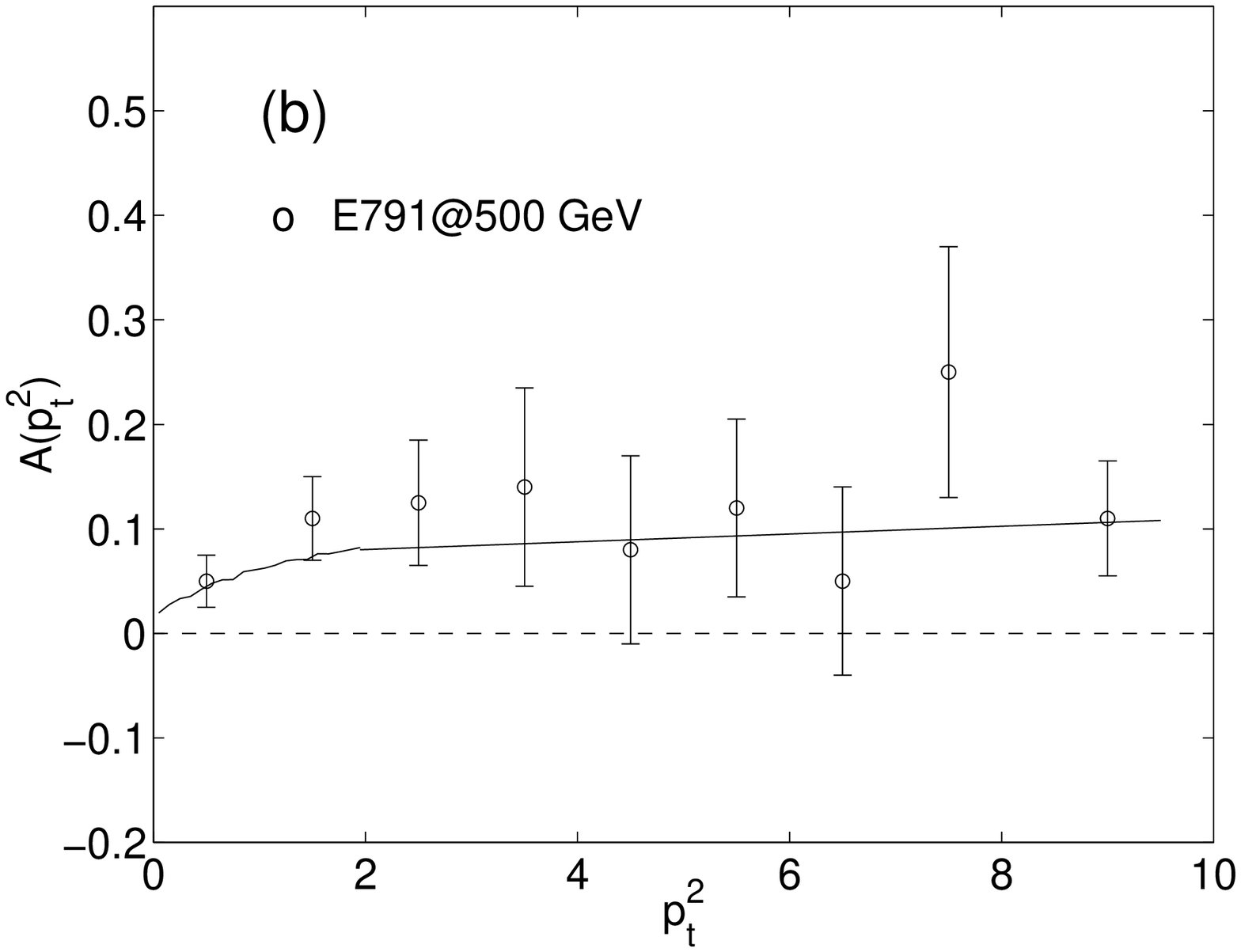,width=79mm}}
\end{center}
\caption[]{The $\D^{\pm}$ asymmetry as a function of (a) $\xF$ and (b) $\pt^2$
for the modified version of \Py~described in the text: new cluster collapse
mechanism, constituent masses for the quarks ($m_c = 1.5$ GeV, $m_d = m_u = 0.33$
GeV), intermediate BRDF's, primordial $\kt = 1.0$ and using two attempts to form two
particles from a cluster. (a) is for all $\pt$ and
(b) is for $-0.2 \leq \xF \leq 0.8$. Experimental data is from \cite{WA82,E769,E791}.}
\label{fig.modified}
\end{figure}

The E791 collaboration has produced a tuned version of \Py~\cite{E791}.
The specific parameters involved have already been discussed above,
so we here only provide a brief summary: 
\begin{Itemize}
\item The charm quark mass is increased from 1.35 to 1.7 GeV. 
The fraction of cluster collapses is thereby reduced by about 35 \%, 
which reduces the  asymmetry considerably for $\xF>0$, as demonstrated
in Fig.~\ref{fig.collapse}. This large charm mass may appear unrealistic but
it is also possible to reduce the number
of cluster collapses by increasing the light-quark masses and/or increasing
the probability for a cluster with a mass above the
$\D\pi$ threshold to decay into two particles. By using constituent masses
both for the light quarks and the charm quark (e.g. $m_{\c} = 1.5$ GeV and
$m_{\d} = m_{u} = 0.33$ GeV) and by allowing two attempts to form two hadrons
from a cluster, the number of collapses is reduced by a comparable amount.
A charm mass of 1.5 GeV is also suggested from calculations of total cross
sections for charm production in next-to-leading order QCD \cite{pertcharm2},
and given the theoretical uncertainty regarding quark masses this value
is not unreasonable.
\item A somewhat less peaked BRDF, like our intermediate scenario.
The asymmetry is reduced somewhat in the region $0< \xF <0.4$, Fig.~\ref{fig.brdf}.
A photoproduction experiment \cite{photoprod} also seems to indicate flat BRDF's, but
these are not determined by any basic principles, and are poorly known,
so further studies will be needed here.
\item The width of the Gaussian primordial $\kt$ distribution is 
increased from 0.44 to 1.0 GeV. As we have noted above, such a number is
unexpectedly large, but in agreement with other data, and therefore rather
standard these days. This allows cluster collapses 
between a charm quark and a beam remnant to occur also at fairly large 
$\pt$ values, thus leading to an essentially $\pt$-independent
asymmetry. Additionally, the $\pt$ kick added to charm quarks and beam remnants tend to
increase the average invariant mass of the produced clusters, thereby
reducing the number of cluster collapses.
\end{Itemize}
Taken together, these parameter changes gives a good fit to data, both 
for the asymmetry and for the shape of the individual charm meson
spectra. Fig.~\ref{fig.modified} shows the asymmetry using default \Py~with
the following modifications: new cluster collapse mechanism, constituent quark
masses, intermediate BRDF's, an increased intrinsic $\kt$ and with two attempts
to form two hadrons from a cluster. The amount of cluster collapse is here
reduced by about a factor of three compared with the default, to $\sim 16$\%.

In summary, we have in this letter described the string fragmentation
approach to charm production in hadronic collisions. A number of
uncertainties have been identified and studied in detail, in particular
the transition from a continuous string-mass distribution to a
discrete hadron-mass one. The conclusion is that the model can 
describe much of the existing data, but also that these data do not
fully constrain the choice of model parameters. Further data on
charm production in $\pi^-\p$ collisions may provide further 
information, as may charm production e.g. in $\e\p$ collisions.
Applications include, among other aspects, bottom production in
hadron colliders where the asymmetry between
$\mathrm{B}$ and $\Bbar$ mesons has to be understood in order to
facilitate a study of CP violation effects \cite{Basym}.
We intend to return to these and other related topics.

\end{document}